\RequirePackage{fix-cm}
\documentclass{svjour3}                                

\usepackage{graphicx}
\usepackage{gensymb}
\usepackage{xcolor}
\usepackage{amsmath,amssymb,amsfonts,float,yhmath}
\begin{document}

\title{Application and reduction of a nonlinear hyperelastic  wall model capturing ex vivo relationships between fluid pressure, area and wall thickness in normal and hypertensive murine left pulmonary arteries
\thanks{Supported in part by the US National Science Foundation (DMS-1615820, DMS-1638521) and by U.K.~Research and Innovation (EPSRC EP/N014642/1, EP/S030875/1, EP/T017899/1), and a Leverhulme Research Fellowship (NAH).}
}
%



\author{Mansoor A.~Haider        \and
       Katherine J. Pearce \and Naomi C.~Chesler \and Nicholas A.~Hill \and Mette S.~Olufsen}


\institute{Mansoor A.~Haider (corresponding author) \at
              Department of Mathematics, North Carolina State University, Box 8205, Raleigh, NC 27695-8205 USA  \\
              \email{mahaider@ncsu.edu}, https://orcid.org/0000-0002-3096-1203          
          \and
           Katherine J.~Pearce \at
              Department of Mathematics, North Carolina State University, Box 8205, Raleigh, NC 27695-8205 USA
           \and
            Naomi  C.~Chesler \at
              Edwards Lifesciences Foundation Cardiovascular Innovation and Research Center \& Department of Biomedical Engineering, University of California, Irvine (UCI), Irvine, CA 92697
           \and 
            Nicholas  A.~Hill \at
            School of Mathematics and Statistics, University of Glasgow, Glasgow, G12 8QQ, U.K.
            \and
           Mette S.~Olufsen \at
              Department of Mathematics, North Carolina State University, Box 8205, Raleigh, NC 27695-8205 USA              
}


\maketitle

\begin{abstract}
Pulmonary hypertension is a cardiovascular disorder manifested by elevated arterial blood pressure together with vessel wall stiffening and thickening due to alterations in collagen, elastin and smooth muscle cells. Hypoxia-induced (type 3) pulmonary hypertension can be studied in animals exposed to a low oxygen environment for prolonged time periods leading to biomechanical alterations in vessel wall structure. This study formulates and systematically reduces a nonlinear elastic structural wall model for a large pulmonary artery, generating a novel pressure-area relation capturing remodeling in type 3 pulmonary hypertension.  The model is calibrated using {\em ex vivo} measurements of vessel diameter  and wall thickness changes, under controlled flow conditions, in left pulmonary arteries isolated from control and hypertensive mice. A two-layer, hyperelastic, anisotropic model incorporating residual stresses is formulated using the Holzapfel-Gasser-Ogden model. Complex relations predicting vessel area and wall thickness with increasing blood pressure are derived and calibrated using the data. Sensitivity analysis, parameter estimation and subset selection are used to systematically reduce the 16-parameter model to one in which a much smaller subset of identifiable parameters is estimated via solution of an inverse problem. Our final reduced model includes a single set of three elastic moduli. Estimated ranges of these parameters demonstrate that nonlinear stiffening is dominated by elastin in the control animals and by collagen in the hypertensive group.  The novel pressure-area relation developed in this study has potential impact on one-dimensional fluids network models of vessel wall remodeling in the presence of cardiovascular disease. 

\end{abstract}

\section{Introduction}
\label{intro}
Pulmonary hypertension (PH), encompassing several cardiovascular disorders and manifested by a mean  pulmonary arterial blood pressure (BP)  above 20 mmHg, is commonly classified into five disease groups \cite{MAB09,SMC19}.  One of these, group 3:``pulmonary hypertension due to lung disease'' includes patients with PH induced by hypoxia (HPH). This disease type can be studied in mice with PH induced  by placing the animals in a low oxygen (hypoxic)  environment.  The response of the cardiovascular system is stiffening and thickening of the pulmonary arteries accompanied by an increase in BP to PH levels. The aim of this study is to use mathematical modeling to devise a relationship between BP and vessel lumen area that characterizes the  structural remodeling of the underlying tissues. For this PH group, vascular remodeling typically starts in the small arteries,  proceeding to the large arteries as the disease advances \cite{M04,RR08}. The arterial wall comprises three layers, the intima, a single layer of endothelial cells, the media which contains large amounts of elastin and smooth muscle cells, and the adventitia mainly composed of collagen (Fig.~\ref{fig:configs}a). In animal models of group 3 PH, vessels stiffen largely due to collagen accumulation \cite{OWT09,WC12} and smooth muscle cell proliferation is known to increase the thickness of the vessel wall \cite{WFM95}.

One advantage of characterizing how PH impacts the pressure-area relationship is that the resulting model can be incorporated into one-dimensional (1D)  fluid dynamics network models  used extensively to study hemodynamics in both systemic  \cite{AP07,BBG16,CGL16,MAP07,O99,VS11} and pulmonary \cite{CQR21,PCO20,QVS14} arteries. 1D fluid dynamics models are especially well suited to  predict flow distribution along the network and wave-propagation, but accurate predictions require  appropriate specification of the pressure-area interaction. Moreover, 1D models can be readily calibrated to {\em in vivo} geometry, flow and/or BP measurements. The 1D fluid dynamics models are derived from the Navier-Stokes equations combined with a state equation relating blood pressure and vessel area, often formulated using an empirical or simple elastic wall model.  These simpler models have the advantage of being specified using a small number of parameters \cite{LWG84,O99,QCP18,VBZ11}, but the disadvantage that the manner in which  tissue remodeling associated with disease  translates to the model is unclear.  While complex tissue mechanics models exist \cite{HGO00,RH19,ZMZ18}, they have not been integrated with 1D fluid dynamics models. One  state-of-the-art tissue mechanics model is the two-layer nonlinear hyperelastic model developed by Holzapfel, Gasser, and Ogden \cite{HGO00} (HGO model) that captures {\em ex vivo} biomechanical deformation of the vessel wall. While this model is complex, it includes parameters that more directly and realistically represent structural elements and constituents within the underlying biological soft tissues.  

In this study, we formulate and systematically reduce a nonlinear hyperelastic structural wall model for the large pulmonary arteries, generating a novel pressure-area relation that can characterize remodeling in HPH.  The model will be calibrated to  {\em ex vivo} biomechanical deformation and wall thickness measurements from control and hypertensive mice.  To do so, we start by formulating a two-layer, anisotropic vessel wall model  using the HGO model formulation \cite{HGO00}. In addition to anisotropy and multiple layers, this model accounts for residual stresses, known to be significant in large pulmonary arteries as evidenced by a large opening angle arising when rings from excised vessels are cut. The rings are obtained from cutting ``a slice" normal to the axial direction, and the opening angle is determined from a radial cut through the ring's circumference \cite{HSD01,XPM08}.  Based on this approach, complex relations determining the dependence of vessel area and wall thickness on blood pressure (BP) are derived.  Our model will be calibrated to  {\em ex vivo} measurements of vessel diameter and wall thickness as a function of pressure in the left pulmonary artery (LPA) in control (CTL) and hypertensive (HPH) mice \cite{TC10}. Since our model is complex, containing 16 parameters, calibration using data is challenging.  To remedy this problem we use sensitivity analysis and subset selection \cite{BVV99,CBCL09,MXP11,QM09} to identify the simplest model and a set of sensitive and identifiable parameters that can be estimated using the model and available data.

\section{Models and Methods}\label{model}

In 1D  cardiovascular fluid dynamics network models, the dependent variables are the transmural blood pressure $p(z,t)$ (mmHg) (the difference between blood pressure in the vessel and the surrounding tissue), the vessel lumen area $a(z,t)$ (cm$^2$), and the average flow $q(z,t)$ (cm$^3$/s) through the vessel. $t$ (s) denotes time and $z$ (cm) is the axial coordinate (along the length of the vessel).  The flow is approximated by integrating over the vessel's cross-section, i.e.~$q(z,t)=2\pi\int_0^{r(z,t)} u_z r \, dr$, where $a(z,t)=\pi (r(z,t))^2$, $r(z,t)$ (cm) is the vessel radius, and $u_z$ (cm/s) is the axial component of the fluid velocity. This equation is evaluated by specifying the velocity profile across the lumen \cite{VS11,O99}. 

The 1D fluid dynamics model is derived from an approximation to the Navier-Stokes equations that conserves volume and momentum.  The derivation is achieved by assuming that the flow is incompressible and Newtonian, that the vessels are cylindrical, and that the blood is incompressible, viscous, and homogeneous with constant density $\rho$ (g/cm$^3$) and blood viscosity $\mu$ (g/(cm s)).  Under these assumptions, the 1D fluid dynamics model can be formulated as 
\begin{equation}
\frac{\partial a}{\partial t} + \frac{\partial q}{\partial z}=0,
\frac{\partial q}{\partial t} + \frac{\partial }{\partial z}\left( \frac{q^2}{a} \right) + \frac{a}{\rho} 
\frac{\partial p}{\partial z} = -\frac{2\pi\nu R}{\delta}\frac{q}{a},
\label{eq:fluidsEq}
\end{equation}
where $\nu=\mu/\rho$ (cm$^2$/s) is the Newtonian fluid kinematic viscosity and $\delta$ (cm) is a boundary layer thickness parameter introduced via the velocity profile equation to ensure no-slip at the vessel walls.  The system of equations is closed by introducing a wall model relating the transmural blood pressure $p(z,t)$ and lumen area $a(z,t)$.

This study derives such a relation by treating the wall as a hyperelastic material integrating the two layer model  by Holzapfel, Gasser and Ogden, often referred to as the HGO model  \cite{HGO00}.  This model incorporates nonlinear effects of residual stresses,  anisotropy, material and geometric nonlinearities,  and contributions of key wall constituents (collagen and elastin) within the vessel wall layers.  A schematic of the wall constituents is shown in Fig.~\ref{fig:configs}, and Table \ref{tab:appTable} (in the Appendix) lists the model parameters and their units. 
\begin{figure*}
\begin{center}
 \includegraphics[scale=0.13]{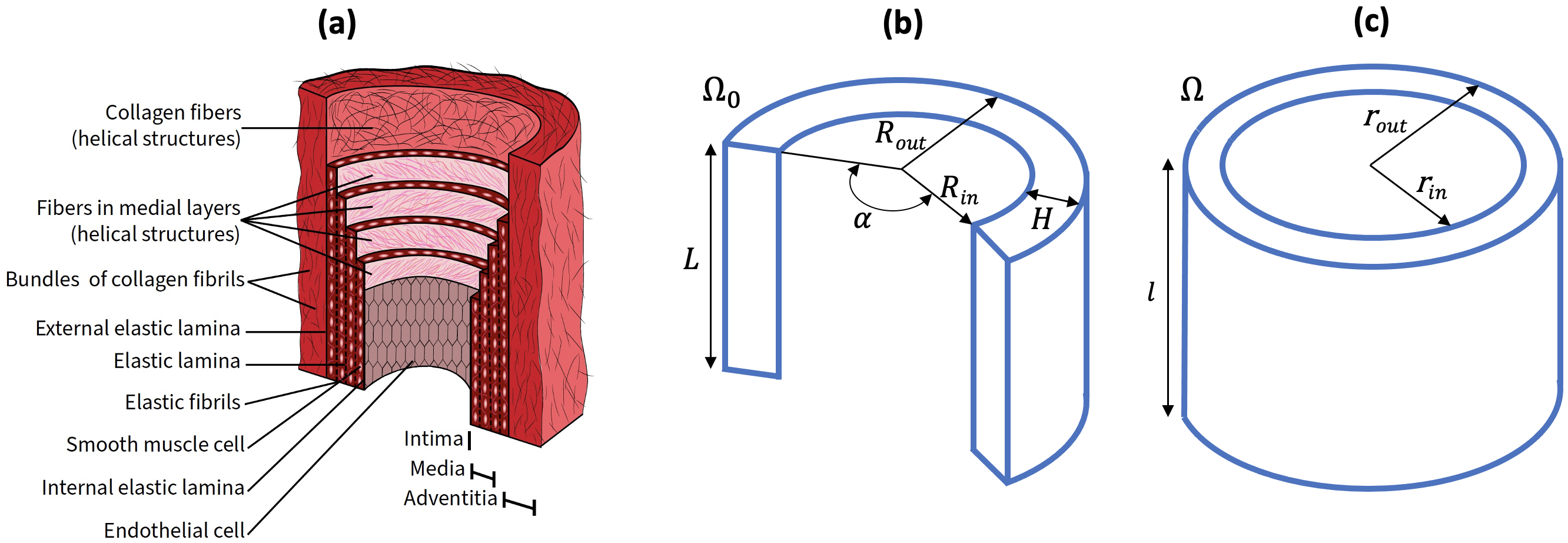}
 \end{center}
\caption{Foundations of the nonlinear hyperelastic wall model: (a) Illustration of a cross-section of a large artery wall (redrawn from \cite{HGO00}); (b) the stress-free reference state $\Omega_0$ defined in equation (\ref{eq:stressFree}) where $R_{in}$ is the inner radius, $R_{out}$ is the outer radius, $H$ is the wall thickness, $L$ is the axial length and $\alpha$ is the opening angle; (c) the  current configuration $\Omega$ defined in (\ref{eq:currentConfig}). Note that the (deformed) inner radius ($r_{in}$), outer radius ($r_{out}$) and  axial length ($l$) are all determined after the model equations are solved. }
\label{fig:configs}  
\end{figure*}

\subsection{Deformation} 
\label{sec:HGOkinematics}

The model is formulated in terms of three configurations of the vessel wall: (i) a  {\em stress-free} reference state $\Omega_0$ (Fig.~\ref{fig:configs}b) represented by  a continuous arc of a cylindrical ring free of all residual stresses; (ii)  an intermediate {\em load-free} configuration (not shown) represented by a closed cylindrical ring  in the absence of fluid flow; and (iii) a  {\em current} configuration $\Omega$ (Fig.~\ref{fig:configs}c) representing the deformed vessel as fluid flows through the vessel lumen in an  {\em ex vivo} or {\em in vivo} setting. 

Specifically, $\Omega_0$ approximates the process of excising a vessel segment, extracting a cross-section approximated as a thin cylindrical ring, and then making a single radial cut along the ring's circumference.  It is denoted by
\begin{equation}
\Omega_0=\left\{ (R,\Theta,Z)\in[R_{in},R_{out}]\times [0,2\pi-\alpha]\times[0,L]
\right\},
\label{eq:stressFree}
\end{equation}
where $(R,\Theta,Z)$ are Lagragian cylindrical (polar) coordinates, $\alpha$ is the opening angle, $L$ is the reference axial length, and $R_{in}$ and $R_{out}$ are the inner and outer radii, respectively.  

Similarly, the current configuration $\Omega$ (shown in Fig.~\ref{fig:configs}c), associated with the deformed vessel representing the coupled state under fluid flow and pressure, is defined as 
\begin{equation}
\Omega=\left\{ (r,\theta,z)\in[r_{in},r_{out}]\times[0,2\pi]\times[0,l]\right\},
\label{eq:currentConfig}
\end{equation}
where  the deformation determines the (unknown) inner radius ($r_{in}$), the outer radius ($r_{out}$), and the vessel length ($l$). 

Finally,  isochoric deformation arising from a state combined inflation, axial extension, and torsion within an elastic tube is denoted by
\begin{equation}
(r,\theta,z) = \left( \sqrt{\frac{R^2 - R_{in}^2}{k\lambda_z}+r_{in}^2},k\Theta + Z\frac{\Phi}{L},\lambda_z Z \right),
\label{eq:deformation}
\end{equation}
where $k=\frac{2\pi}{2\pi-\alpha}$, $\lambda_z$ is the (constant) axial stretch  and $\Phi$ is the twist angle. 
Note that, in  {\em ex vivo} studies with coupled flow and deformation, the parameter $\lambda_z$ should be set based on the observed ratio of the axial length of a vessel segment before and after excision, noting that $\lambda_z>1$ due to residual stresses  {\em in vivo}.  For the data used in this study, $\lambda_z =1.4$ for both the control and hypertensive animals, as reported in \cite{TC10}. The {\em ex vivo} measurements of deformation, after introducing fluid flow through the vessel, are performed in vessels stretched and mounted to match this measured ratio.

\subsection{Two-layer hyperelastic model}
\label{sec:HGOlayers}
Within the HGO  framework, a two-layer hyperelastic wall model accounting for the media ($\gamma=M$) and adventitia ($\gamma=A$) (Fig.~\ref{fig:configs}a) is formulated by representing the Cauchy stress $\sigma=\sigma_M+\sigma_A$ as the sum of the  stress in each layer \cite{HGO00},
\[
\sigma_\gamma=c_\gamma \mbox{dev}\left(J^{-2/3}{\bf b} \right) + 2 \frac{\partial \Psi_\gamma}{\partial \tilde{I}_{1\gamma}}\mbox{dev}\left({\bf a}_{1\gamma}\otimes {\bf a}_{1\gamma} \right) 
\]
\begin{equation} 
\;\;\;\;\;\;\;\;\;\;\;
+ 2 \frac{\partial \Psi_\gamma}{\partial \tilde{I}_{2\gamma}}\mbox{dev}\left({\bf a}_{2\gamma}\otimes {\bf a}_{2\gamma} \right),\;\;\gamma=M,A,
\label{eq:stressStrain}
\end{equation}
where $\Psi_\gamma$ is the Helmholtz free energy for each layer, and has the form,
\begin{equation}
\Psi_\gamma=\frac{k_{1\gamma}}{2k_{2\gamma}}\left[ e^{k_{2\gamma}(\tilde{I}_{1\gamma}-1)^2} + e^{k_{2\gamma}(\tilde{I}_{2\gamma}-1)^2} - 2 \right],\gamma=M,A
\label{eq:SEDef}
\end{equation}
In (\ref{eq:stressStrain}), $c_\gamma$ represent the elastic moduli for the isotropic constituents (mostly elastin) in each layer, $J=\mbox{det}({\bf F}),$ where ${\bf F}$ is the deformation gradient of (\ref{eq:deformation}), ${\bf b}={\bf F}{\bf F}^T$,  and $\tilde{I}_{l\gamma} = {\bf A}_{l\gamma}:\bar{\bf C}$ where $\bar{\bf C}=J^{-2/3}{\bf C}$ (${\bf C}={\bf F}^T{\bf F}$) and ${\bf A}_{l\gamma}={\bf a}_{0l\gamma}\otimes{\bf a}_{0l\gamma} (l=1,2,\;\gamma=A,M)$.   In (\ref{eq:SEDef}), $k_{1\gamma}$ and $k_{2\gamma}$ are elastic parameters for the anisotropic constituents (mostly collagen) in each layer (Fig.~\ref{fig:configs}a).  Lastly, Eulerian and Lagrangian vectors, ${\bf a}_{l\gamma}$ and ${\bf a}_{0l\gamma}$ (respectively),   associated with collagen fiber directions are determined via ($\gamma=A,M$),
\begin{equation}
{\bf a}_{l\gamma}=J^{-1/3}{\bf F}{\bf a}_{0l\gamma}, \;{\bf a}_{0l\gamma}=\left( \begin{array}{c} 0 \\ \;\;\;\cos(\beta_{\gamma}) \\  \pm\sin(\beta_{\gamma})\end{array}\right), l=1,2,
\label{eq:fiberVecs}
\end{equation}
where $\beta_{\gamma}$ are the  fixed collagen fiber angles in each layer (Fig.~\ref{fig:configs}a).

\subsection{Pressure-area relation}
\label{sec:pa}

We obtain a hyperelastic pressure-area relation by integrating the radial component of the stress equilibrium equation. Neglecting inertial terms and assuming a quasi-static state this stress equilibrium equation, expressed in the current configuration, is given by
\begin{equation}
\frac{d\sigma_{rr}}{d r} + \frac{\sigma_{rr}-\sigma_{\theta\theta}}{r}=0,\;\;r_{in}<r<r_{out},
\label{eq:stressEq}
\end{equation}
where $r_{in}=r(R_{in})$ and $r_{out}=r(R_{in}+H)$, and $\sigma_{rr}, \sigma_{\theta\theta}$ are the radial and circumferential normal stress components, respectively.  Here, $H$  denotes the undeformed vessel wall thickness (Fig.~\ref{fig:configs}b).

Balance of forces between the transmural blood pressure and the radial component of the normal stress in the wall is enforced by the condition,
\begin{equation}
p=\left.-\sigma_{rr}\right|_{r=r_{in}}\Rightarrow p=\int_{r_{in}}^{r_{out}}\frac{\sigma_{rr}-\sigma_{\theta\theta}}{r} dr.
\label{eq:pressBC}
\end{equation}
Equations (\ref{eq:deformation}-\ref{eq:stressStrain}) are used to formulate the integrand in (\ref{eq:pressBC}), which is evaluated with the aid of symbolic computation software (MAPLE 2019). 

The resulting pressure-area relation can be written as
\[
p=\int_{r_{in}}^{r_{MA}} {\cal F}_M(r_{in},r)dr + \int_{r_{MA}}^{r_{out}} {\cal F}_A(r_{in},r)dr,
\]
\begin{equation}
\;\;\;\;\;\;\;\;\;\;\;\;\;\;\;\;\;\;\;\;\;\;\;\mbox{where}
\;\;r_{MA}=r(R_{in}+H_{M}),
\label{eq:pAInt}
\end{equation}
and $H_{M}$ is the (reference) thickness of the media.  
Recall that the relation  $r_{in}=\sqrt{a(z,t)/\pi}$ is used to express the inner radius (\ref{eq:pAInt}) in terms of  the vessel  area, ultimately for incorporation into  (\ref{eq:fluidsEq}).  For brevity, the mathematical forms of the integrands ${\cal F}_M$ and ${\cal F}_A$ are not included here as these are lengthy expressions imported from MAPLE into MATLAB (R2021b). The integral is evaluated numerically using the MATLAB ``integral" command which employs global adaptive quadrature \cite{S08} (see note in Appendix, \S5.2). 

This final pressure-area relation  \eqref{eq:pAInt} contains 16 model parameters
\begin{equation*}
  {\bf q} = [ R_{in},R_{out},H,H_{M},\alpha,L,\Phi,\lambda_z,...
\end{equation*}
\begin{equation}
\;\;\;\;\;\;\;\;\;\;\;\;\;\;\;\;\;\;\;\;\;c_M, k_{1M}, k_{2M},\beta_M, c_A, k_{1A}, k_{2A},\beta_A],\\
\end{equation}
listed with units and values in  the Appendix (Table \ref{tab:appTable}). For a fixed set of these parameters, the model prediction of wall thickness is evaluated using equations (\ref{eq:pAInt}) and  (\ref{eq:deformation}) via the difference $r(R_{in} + H)-r(R_{in})$.

\subsection{{\em Ex vivo} murine data}
\label{sec:data}

The model developed above is calibrated to murine data made available by Chesler. The majority of data along with detailed descriptions of the experiments can be found in the study by Tabima and Chesler \cite{TC10}.

Data measuring  lumen area and wall thickness changes with increasing transmural blood pressure were measured  under  {\em ex vivo} biomechanical testing in excised left pulmonary artery (LPA) vessel segments from male C57BL6 mice under control (CTL) and 10-day hypoxia-induced hypertensive (HPH) conditions \cite{TC10}. In both the control (CTL) and hypertensive (HPH) vessel segments, 11  measurements ($i=1,\ldots,11$) are included for the relationship between vessel outer diameter ($D_i^{\text{data}}$) and pressure ($p_i^{\text{data}}$), and 3  measurements ($j=1,2,3$)  for the relationship between vessel wall thickness ($T_j^{\text{data}}$) and pressure ($p_j^{\text{data}}$). For each group, these measurements represent average values over 4 control (CTL) and 5 hypertensive (HPH) animals under controlled flow conditions with pressures in the range of 0-50 mmHg. Specific pressure values for each group are noted in Fig.~\ref{fig:8param}a-b. 

\subsection{Fixed model parameters} 
\label{sec:asssump}
Several of our model parameters can be fixed at representative values based on literature values or details of the experiments used to calibrate the models.  First, we assume that the vessels have no  twist, i.e.~$\Phi=0\degree$ and  that the opening angle in the stress-free reference state is $\alpha=94.2\degree$. This latter value was obtained from literature reporting measurements in rings extracted from murine LPA vessels \cite{XPM08}. 
Moreover, to mimic the {\em in vivo} setting, excised vessels were stretched to 140\% of their length after extraction prior to mechanical testing \cite{TC10}, corresponding to a fixed value of $\lambda_z=1.4$  in (\ref{eq:deformation}). Since detailed histology for the murine LPA is not available, we use literature values reported in the HGO model \cite{HGO00} to set collagen fiber angles  at $\beta_{M}=29\degree$ and $\beta_A=62\degree$.   Finally, we assume that the media occupies 2/3 of the vessel wall thickness in the stress-free reference state.  These fixed parameter values are summarized in the Appendix (Table \ref{tab:appTable}). Accounting for these six assumptions, for the parameter dependency $R_{out}=R_{in}+H$ and observing that the model is independent of $L$ yields the following 8 parameters to be estimated 
\begin{equation}
  {\bf q}_8 = [ R_{in},H, c_M, k_{1M}, k_{2M},c_A, k_{1A}, k_{2A}].
  \label{eq:8paramvec}
\end{equation}

\subsection{Parameter estimation, sensitivity, identifiability and model reduction}
\label{sec:PEI}

In the context of our model and data, we formulate a parameter estimation problem determining $m$ parameters ${\bf q}^*$ that minimize the least squares cost ${\cal J}$ as
\begin{equation}
{\bf q}^* =  \arg \min_{{\bf q} \in \mathbb{R}^{m}_{\geq 0}} {\cal J}({\bf q}),\;\;\mbox{where:}\;\;{\cal J}({\bf q})={\bf s}({\bf q})^T {\bf s}({\bf q}),
\label{eq:cost}
\end{equation}
where the 14-component residual vector ${\bf s}({\bf q})$  is given by
\begin{eqnarray}
   {\bf s}({\bf q}) &=& \left[  {\bf s}_1, {\bf s}_2 \right], \label{eq:residual} \\
   {\bf s}_1 &=& \frac{1}{\sqrt{n_1}}  \left( \frac{ p(a^{\text{data}}_{\text{in},i}) - p^{\text{data}}_i}{p^*}\right) \nonumber \\
   {\bf s}_2 &=& \frac{1}{\sqrt{n_2}}  \left( \frac{T(p^{\text{data}}_{j}) - T^{\text{data}}_j}{T_1^{\text{data}}} \right) \nonumber
\end{eqnarray}
with  $i=1,\ldots,n_1$ and $j=1,\ldots,n_2$, with $n_1=11$ and $n_2=3$.  

The optimization problem is solved using a Nelder-Mead direct search simplex algorithm \cite{NM65} minimizing ${\cal J}({\bf q})$ in (\ref{eq:cost}) using the routine ``fminsearch" in Matlab.

Note, the mathematical model is used to evaluate the term  $a^{\text{data}}_{\text{in},i}$  by first converting the outer diameter data ($D_i^{\text{data}}$) to an inner radius using (\ref{eq:deformation}) and then using (\ref{eq:pAInt}) to determine the values $ p(a^{\text{data}}_{\text{in},i})$.  The term $T(p^{data}_j)$ is evaluated as outlined at the end of \S\ref{sec:pa}.   Calibration of the model to data is done in an iterative manner, gradually reducing the model complexity and number of parameters estimated using sensitivity analysis and  subset selection.

\paragraph{Sensitivity analysis} is performed after parameter estimation (with $n$ data points) using local methods calculating the $n\times m$ sensitivity matrix $\chi = \nabla_{\bf q}  {\bf s}({\bf q})$ using a first order finite difference scheme. Prior to calculation of sensitivity derivatives, a linear mapping is used to normalize across scales due to the  diverse set of parameters and units in our model.  Specifically, a perturbed interval about the $k$th parameter estimate $[(1-\alpha)q_k^*,(1+\alpha)q_k^*]$ is mapped to $[0,1]$ via the linear transformation $y=\frac{1}{2\alpha}(\alpha -1 +\frac{x}{q_k^*})$. This yields, via the Chain rule, the derivative transformation $\frac{\partial (\cdot)}{\partial x}=\frac{\partial (\cdot)}{\partial y}\frac{dy}{dx}$, resulting in a multiplying factor of $2\alpha q_k^*$ in transforming raw sensitivities to their scaled counterparts.  The value $\alpha=0.1$ was prescribed and all sensitivity derivatives above were approximated using first-order finite difference approximations with a step size chosen sufficiently small ($10^{-7}$). This choice ensured numerical convergence of all scaled parameter sensitivity derivative computations across all cases considered in this study. 

\paragraph{Subset selection and model reduction} is performed using the eigenvalue method \cite{BVV99,CBCL09,MXP11,QM09}  that analyzes the magnitude of eigenvalues and corresponding eigenvectors for the  $m\times m$  Fisher information matrix approximated as $\chi^T \chi$ at ${\bf q}={\bf q}^*$  \cite{R71}. In our study, the eigenvalue subset selection method is guided by physical properties of our model at each stage of the overall process.  At  ${\bf q}={\bf q}^*$, our subset selection  analysis and model reduction is summarized by the following iterative procedure.

\begin{enumerate}
\item Determine the eigenvalues of the matrix $\chi^T \chi$.
\item Check if the smallest eigenvalue of $\chi^T \chi$ is below a specified threshold $\eta$ (see, e.g.~Fig.~\ref{fig:8paramPSS}). 
\item If step 2 is satisfied, examine the eigenvector corresponding to the smallest eigenvalue.
\item Mark the order 1 components of the eigenvector in step 3.
\item Parameters corresponding to the marked vector components in step 4 are potentially {\em unidentifiable} and considered as candidates for fixing at nominal values, or uncovering parameter dependencies.
\item  If possible, we  reduce the model by fixing or eliminating unidentifiable parameters.
\end{enumerate}
During the course of this iterative procedure, we ensure that the cost ${\cal J}({\bf q}^*)$ is preserved.  This approach strikes a balance between model reduction and robust optimization,  preserving  the quality of curve-fits within the context of the given data set as the process advances.

\section{Results}
We apply the iterative approach for estimating the non-fixed parameters in (\ref{eq:8paramvec}) as outlined below. 

\begin{itemize}
\item[\S\ref{sec:8param}] estimates the 8 non-fixed parameters for the control animals. Results of parameter estimation, sensitivity analysis, and subset selection yield a reduced model with 6 parameters.
\item[\S\ref{sec:6param}] estimates these 6 parameters for the control animals using the reduced model.  Results of this analysis enabled further model reduction, yielding a model with equal elastic moduli in the two layers. The resulting model has 5 parameters.
\item[\S\ref{sec:5param}] estimates these 5 parameters in the further reduced model for both control and hypertensive animals.  Results reveal that two parameters are correlated. Fixing one of the correlated parameters yields the final 4-parameter model. 
\item[\S\ref{sec:rangeEst}] examines the final, fully reduced  4-parameter model. To investigate effects of fixing one of the two correlated parameters identified in \S\ref{sec:5param}, we conducted a study evaluating impacts of varying the fixed parameter. We report  parameter ranges for successful results,  preserving quality of curve-fits via bounds on the least squares error for both control and hypertensive animals.
\end{itemize}

\begin{figure}
\begin{center}
 \includegraphics[scale=0.285]{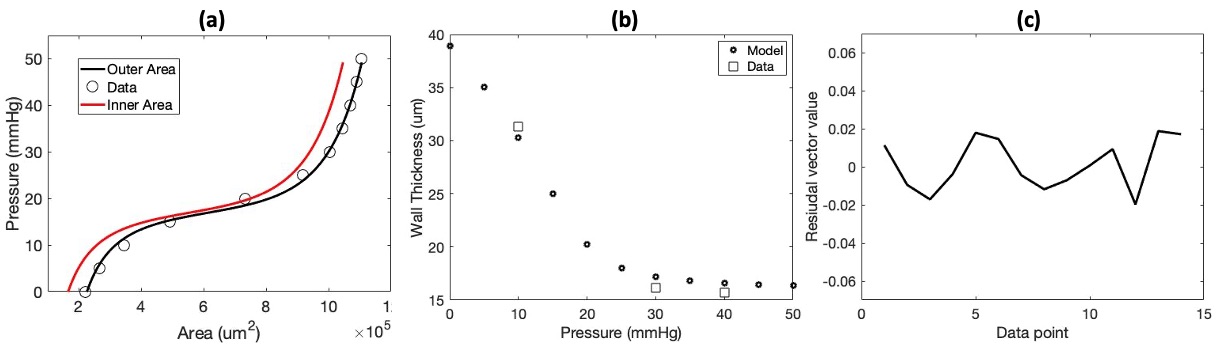}
 \end{center}
\caption{Results from estimating 8 parameters (listed in Table \ref{tab:CTLparamFit}) for the control (CTL)  animals.  (a) pressure vs.~area, model predictions of the outer area (black) vs.~data (circles) and the inner area (red); (b) wall thickness vs.~pressure model predictions compared to the 3 data points (squares); (c)  the residual vector (\ref{eq:residual}) across the 14 data points.}
\label{fig:8param}       
\end{figure}

\subsection{Baseline control animal model (8 parameters)}
\label{sec:8param}

We first estimate the 8 non-fixed parameters for the control animals
\begin{equation}
  {\bf q}_8 = [ R_{in},H, c_M, k_{1M}, k_{2M},c_A, k_{1A}, k_{2A}].
  \label{eq:8paramvec2}
\end{equation}
Initial and estimated parameter values for this case are reported in Table \ref{tab:CTLparamFit} (initial values are also shown Table \ref{tab:appTable} in the Appendix). 

The initial value of the reference wall thickness $H$ is set based on the first data measurement in the experiments. The initial  value of $R_{in}$ ($1$ mm) is set using an order of magnitude estimate for the LPA.  Setting  initial values for the remaining 6 parameters is challenging given that experiments yielding direct measurements for their (nominal) values do not exist.  Consequently, isotropic elastic moduli $c_M, c_A$ are  set to initial values of $10$ kPa, an accurate order of magnitude  estimate for this type of biological soft tissue.  Due to nonlinearity in the anisotropic portions of the model, initial values for the remaining parameters are determined by systematic variation of initial parameter choices.  Initial values for these parameters yielding a high cost ${\cal J}$ were rejected to arrive at at a combination of initial values with a curve fit to the data of good quality. The resulting combination of initial values for  $ k_{1M}$ and $ k_{1A}$ is 1 kPa and   0.3 kPa, respectively, while the  initial values for $k_{2M}$ and $k_{2A}$ are based on those reported in the HGO study \cite{HGO00}.  This combination of initial values yields the most consistent set of results across all cases considered. 
 
Model predictions with estimated parameters depicting  pressure vs.~area and the wall thickness vs.~pressure (shown in Fig.~\ref{fig:8param}a and b) provide excellent fits to the control animals. Inspection of estimated parameters reveal that the adventitia parameter value $k_{2A}\approx 0$ (see Table \ref{tab:CTLparamFit}), suggesting that we can eliminate the  anisotropic terms for the adventitia (the last two terms in (\ref{eq:stressStrain})). This observation suggests  that the parameter $k_{1A}$ is  structurally unidentifiable since this parameter only appears in the last two (anisotropic) terms of (\ref{eq:SEDef}) in the adventitia ($\gamma=A$). 

\begin{figure*}
\begin{center}
 \includegraphics[scale=0.36]{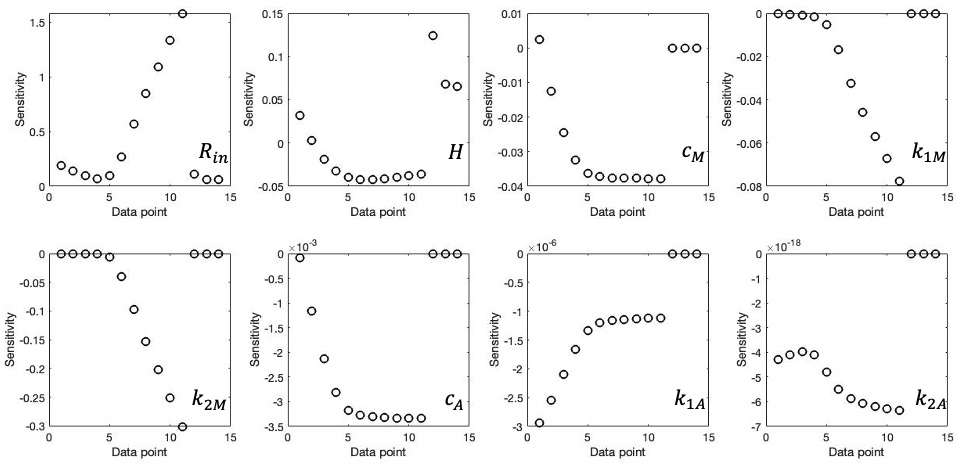}
 \end{center}
\caption{Normalized parameter sensitivities  for the control  (CTL) animals with 8 estimated parameters across the 14 data points.}
\label{fig:8paramSens}
\end{figure*}

\begin{figure*}
\begin{center}
 \includegraphics[scale=0.35]{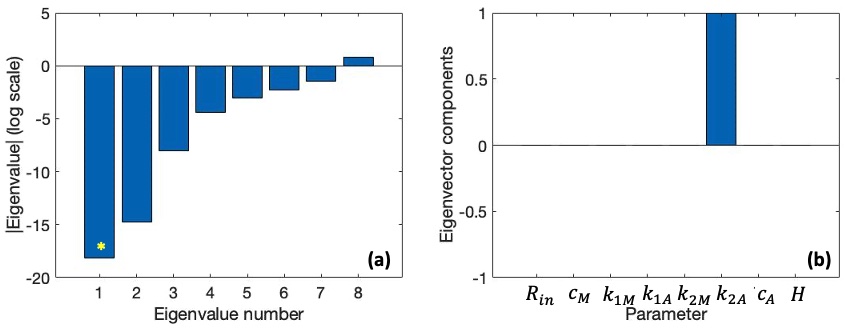}
 \end{center}
\caption{Identifiability results using the eigendecomposition of  the information matrix ($\chi^T\chi$)  for the control (CTL) animals with  8 estimated parameters: (a) log-plot of the eigenvalues of $\chi^T\chi$; (b) components of the  eigenvector of $\chi^T\chi$ corresponding to the smallest eigenvalue of $\chi^T\chi$ that is less than $\eta=10^{-10}$ (asterisk).}
\label{fig:8paramPSS}      
\end{figure*}

The Fisher information matrix $\chi^T \chi$ is used to evaluate the eigenvalues depicted in Fig.~\ref{fig:8paramPSS}a.  Examination of the eigenvector of $\chi^T\chi$ (Fig.~\ref{fig:8paramPSS}c) corresponding to its smallest eigenvalue with $\eta< 10^{-10}$ flags the parameter  $k_{2A}$ (has an order 1 component), indicating that this parameter is unidentifiable.  This designation is consistent with result of sensitivity analysis (shown in Fig.~\ref{fig:8paramSens}), which demonstrates that the sensitivities for both  for $k_{1A}$ and $k_{2A}$  are small relative to the other parameters.

Taken together, these findings motivate a model reduction in which $k_{1A}\approx k_{2A}\approx 0$. Thus, in the next step we analyze a 6-parameter reduced model  eliminating the   anisotropic terms  for the adventitia in the stress-strain law. 

\subsection{Reduced control animal model (6 parameters)}
\label{sec:6param}

\begin{figure*}
\begin{center}
 \includegraphics[scale=0.235]{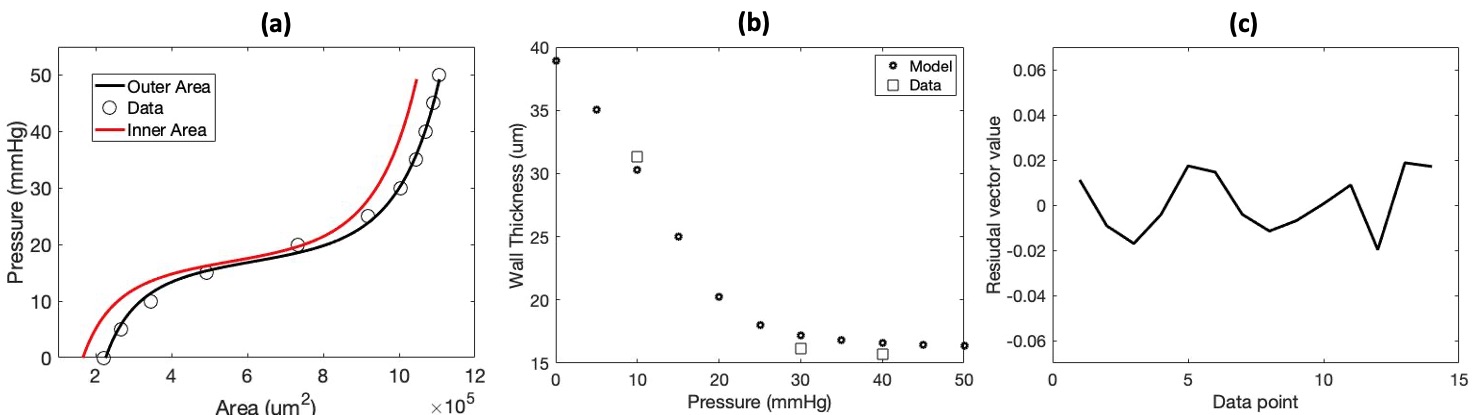}
 \end{center}
\caption{Parameter estimation results for the reduced model for the control  (CTL) animals with 6 estimated parameters: (a) pressure vs.~area model predictions of the outer area (black) vs.~data (circles) and the inner area (red); (b) wall thickness vs.~pressure model predictions compared to the 3 data points (squares); (c)  plot of the  residual vector (\ref{eq:residual}) across the 14 data points.}
\label{fig:6param}      
\end{figure*}

Parameter values are initialized as described in \S\ref{sec:8param}. The 6 parameters to be estimated for the control animals are
\begin{equation}
  {\bf q}_6 = [ R_{in},H, c_M, k_{1M}, k_{2M},c_A].
  \label{eq:6paramvec}
\end{equation}
Results are listed in Table \ref{tab:CTLparamFit}. Again, for the control animals the quality of curve fits of the model to the pressure vs.~area  data (Fig.~\ref{fig:6param}a) and the wall thickness vs.~pressure data (Fig.~\ref{fig:6param}b) is preserved, with a slight reduction in overall cost from ${\cal J}=1.707\cdot 10^{-4}$ to ${\cal J}=1.674\cdot 10^{-4}$. In addition, the estimated values are preserved within $0.25\%$ for the geometric parameters ($R_{in},H$), concurrent with a $3.0\%$ reduction in the elastic modulus $k_{1M}$, a  11.5\% reduction in the elastic modulus $c_M$ and a substantial increase ($2.4\times$) in the modulus $c_A$. Finally, the dimensionless parameter $k_{2M}$ increases by $1.5\%$.

\begin{figure*}
\begin{center}
 \includegraphics[scale=0.32]{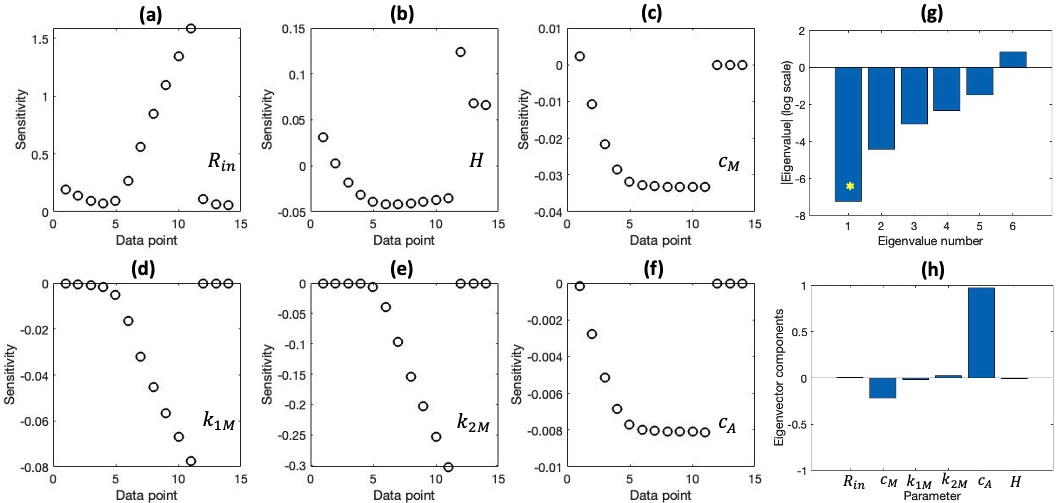}
 \end{center}
\caption{Identifiability results computed using eigendecomposition of  the information matrix ($\chi^T\chi$)  for the reduced 6 parameter model with data from the control (CTR) animals: (a-f) normalized parameter sensitivities for the 6 estimated parameters across the 14 data points; (g) log-plot of the eigenvalues of $\chi^T\chi$; (h) components of the  eigenvector of $\chi^T\chi$ corresponding to the smallest eigenvalue of $\chi^T\chi$ less than $\eta=10^{-7}$ (asterisk).}
\label{fig:6paramPSS}       
\end{figure*}

Examination of the eigenvector of $\chi^T\chi$ (Fig.~\ref{fig:6paramPSS}h) corresponding to its smallest eigenvalue with ($\eta< 10^{-7}$) shown in Fig.~\ref{fig:6paramPSS}g flags the two parameters $c_A$ and $c_M$ with order 1 components; $c_A$ is the  dominant component.  The sensitivity for $c_A$ is also small relative to the other parameters (Fig.~\ref{fig:6paramPSS}a-f). These findings, combined with the observation that $c_M$ and $c_A$ exhibit the largest changes in estimated values between the 8-parameter and 6-parameter fits, motivate a further reduced 5-parameter model examined in the next stage of the process. 

Since two of the remaining 5 parameters are geometric parameters, we retain three elastic parameters describing the isotropic and anisotropic responses in the model.  Thus, the only elastic parameters estimated in the next step are $c_M, k_{1M}$ and $k_{2M}$.  Specifically, the reduced model analyzed in the next section has elastic moduli in the media and adventitia that are set equal, whereas the two layers still retain distinct collagen fiber angles ($\beta_M, \beta_A$).

\begin{table*}[tbh!]
\caption{Estimated parameter values for the control (CTL) animals with the 8-parameter model (column 5) and the reduced 6-parameter model (column 6). }
\centering
\begin{tabular}{|c||c|c|l||l|l||}
\hline
& Param. & Units & Initial & Baseline (\S\ref{sec:8param}) & Reduced (\S\ref{sec:6param})   \\
\hline
$m$ & & & & 8 & 6   \\
\hline
Geom. & $R_{in}$ & $\mu$m & 1000 & 373.809 &  374.654   \\
& $H$ & $\mu$m & $T_1^{data}$&  45.742 & 45.652  \\
\hline
Media & $c_M$ & kPa & 10  & 28.265 & 25.028  \\
& $k_{1M}$ & kPa &  1 & 0.631 & 0.612  \\
& $k_{2M}$ & - & 0.839 &  0.554 & 0.564 \\
\hline
Adv. & $c_A$ & kPa & 10  & 5.392 & 13.162 \\
& $k_{1A}$ & kPa & 0.3 &  0.055 &   0.000 (fixed) \\
& $k_{2A}$ & - &  0.711 & 0.000 &  0.000 (fixed)  \\ 
\hline
${\cal J}$ ($\cdot 10^{-4}$) &  & - & & 1.7071 &1.6742  \\
\hline 
\end{tabular}
\label{tab:CTLparamFit}
\end{table*}

\subsection{Reduced control and hypertensive animal model (5 parameters)}
\label{sec:5param}

In the reduced 5-parameter model, elastic parameters in the two layers are assumed equal, i.e., 
$c_A=c_M$, $k_{1A}=k_{1M}$, and $k_{2A}=k_{2M}$. The parameter vector estimated for this model is
\begin{equation}
  {\bf q}_5 = [ R_{in},H, c_M, k_{1M}, k_{2M}].
  \label{eq:5paramvec}
\end{equation}

This  model is fitted to data from both the control (CTL) and hypertensive (HPH) animals. Values of the five model parameters are initialized as described in \S\ref{sec:8param} and the estimated parameter values  are reported in Table \ref{tab:5paramFit}.  For comparison, results of the 6-parameter model are also included in the table. Note that the estimated parameter values of $c_M, k_{1M},$ and $k_{2M}$ for this reduced model should be interpreted as aggregate elastic parameters for the entire vessel wall, i.e., representing all layers. 

\begin{table*}[tbh!]
\caption{Estimated parameter values for the reduced 5-parameter model for control (CTL) (column 6) and hypertensive (HPH) (column 5) animals.   For comparison, results for the reduced  6-parameter model  are also shown (column 4).}
\centering
\resizebox{\columnwidth}{!}{
\begin{tabular}{|c||c|c|r||r|r||r|}
\hline
& Param. & Units & Initial & CTL (\S 3.2) &  CTL (\S 3.3) & HPH (\S 3.3)   \\
\hline
$m$ & & & & 6 & 5 & 5  \\
\hline
Geom. & $R_{in}$ & $\mu$m & 1000 & 374.654 & 376.632 & 394.114  \\
& $H$ & $\mu$m & $T_1^{data}$&  45.652 & 45.435 & 61.519 \\
\hline
Media & $c_M$ & kPa & 10  & 25.028 & 21.497 & 0.580 \\
& $k_{1M}$ & kPa &  1 & 0.612 & 0.602 & 13.385 \\
& $k_{2M}$ & - & 0.839 &  0.564 & 0.581 & 1.295\\
\hline
Adv. & $c_A$ & kPa & 10  & 13.162 & $=c_M$ &  $=c_M$\\
& $k_{1A}$ & kPa &  &  (fixed at 0.0) &  $=k_{1M}$ &  $=k_{1M}$\\
& $k_{2A}$ & - &   & (fixed at 0.0) &  $=k_{2M}$ &$=k_{2M}$ \\ 
\hline
${\cal J}$ ($\cdot 10^{-4}$) &  & - & & $1.6742$ & $1.6406$ & 0.7785 \\
\hline 
\end{tabular}
}
\label{tab:5paramFit}
\end{table*}

\begin{figure*}
\begin{center}
 \includegraphics[scale=0.28]{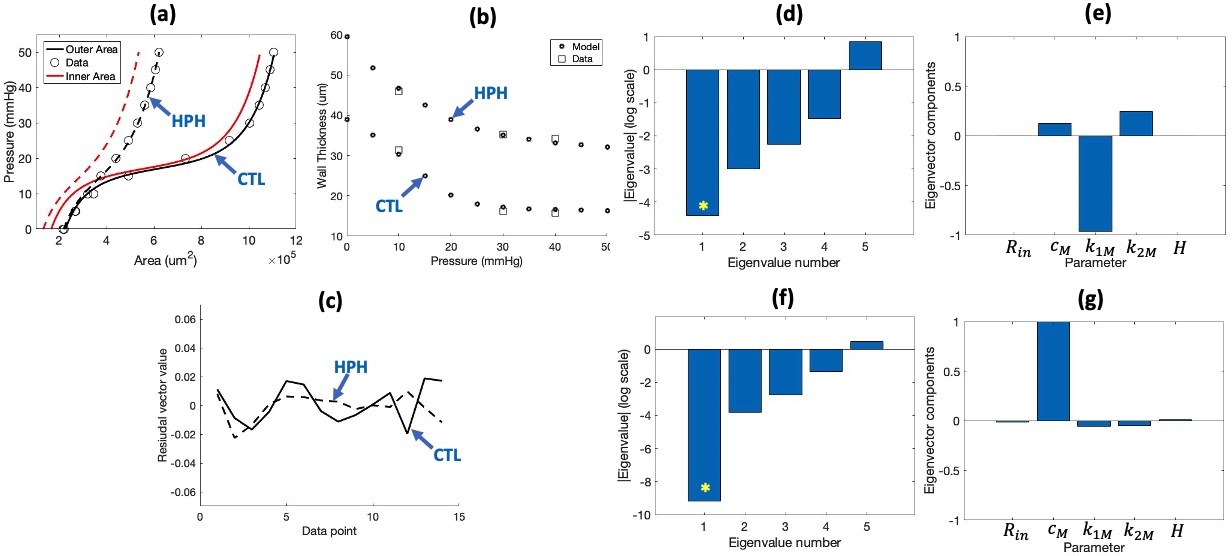}
 \end{center}
\caption{Parameter estimation and identifiability results for the reduced 5-parameter model for the  control (CTL) and hypertensive (HPH) animals: (a) pressure vs.~area model predictions of the outer area (black) vs.~data (circles) and the inner area (red); (b) wall thickness vs.~pressure model predictions compared to the 3 data points (squares); (c)  plot of the  residual vector (\ref{eq:residual}) across the 14 data points; (d,f) log-plot of the eigenvalues of $\chi^T\chi$ in the normotensive (d) and hypertensive (f) animals; (e,g) components of the  eigenvector of $\chi^T\chi$ corresponding to the smallest eigenvalue of $\chi^T\chi$ (asterisk) in the control (e) and hypertensive (g) animals. }
\label{fig:5param} 
\end{figure*}

For the control (CTL) animals, the quality of curve fits of the model to the pressure vs.~area data (Fig.~\ref{fig:5param}a) and the wall thickness vs.~pressure data (Fig.~\ref{fig:5param}b) are preserved, with a slight reduction in overall cost from ${\cal J}=1.674\cdot 10^{-4}$ to ${\cal J}=1.641\cdot 10^{-4}$. The curve fit for hypertensive (HPH) animals has a significantly lower cost (${\cal J}=0.778\cdot 10^{-4}$) due, in part, to the smaller range of variation in the pressure-area curve caused by vessel wall stiffening  (Fig.~\ref{fig:5param}a \& Table \ref{tab:5paramFit}).  In the hypertensive animals, geometric parameters exhibit a small increase in vessel wall inner radius $R_{in}$ ($\approx394\mu$m  vs.~$\approx 377\mu$m) and a large (and expected) increase in reference wall thickness $H$ ($\approx62\mu$m vs.~$\approx45\mu$m).  The altered dynamics in the hypertensive animals are reflected by a substantial increase in the elastic modulus $k_{1M}$ (13.38kPa vs.~0.60kPa) and in the dimensionless parameter $k_{2M}$ (1.30 vs.~0.58), both associated with collagen stiffness.   Concurrently, in the hypertensive animals there is a substantial drop in the elastic modulus $c_M$ (21.50kPa vs.~0.58kPa), associated with elastin deformation, relative to the control animals. 

For this model, subset selection evaluating the eigenvalues and eigenvectors of $\chi^T\chi$   (Fig.~\ref{fig:5param}d-g) for the estimated parameters (Table \ref{tab:5paramFit}, Fig.~\ref{fig:5param}d-g) reveals that the eigenvectors for the smallest eigenvalue  (Fig.~\ref{fig:5param}d,f) flag two parameters $k_{1M}$ and $c_M$  as the dominant components in the control (CTL) and hypertensive (HPH) animals, respectively    (Fig.~\ref{fig:5param}e vs.~Fig.~\ref{fig:5param}g).  These observations suggest a  parameter dependency between the two elastic moduli $c_M$ and $k_{1M}$, which share the same units, that is evaluated in the next section using  our final  4-parameter model.

\subsection{Parameter dependencies and range estimates for the final model (4 parameters)}
\label{sec:rangeEst}

To study the implications of fixing one of the two correlated parameters, for both the control and hypertensive animals, our final model  fixes $c_M$ while still estimating $k_{1M}$, i.e., analysis in this section estimates the 4 parameters,
\begin{equation}
{\bf q_4} = \{  R_{in}, H, k_{1M}, k_{2M}\}.
\end{equation}
Since we do not have data from independent experiments for $k_{1M}$, we repeat optimization while varying this parameter. To preserve quality of the curve fits, parameter ranges were determined by enforcing the condition that the cost increased by no more than 10\% (to two decimal places) relative to the values in Table \ref{tab:5paramFit}, and  across several runs of the optimization. Based on this criteria, the maximum costs used as cutoffs  are set to  ${\cal J}=1.81\cdot 10^{-4}$ (CTL) and ${\cal J}=0.86\cdot 10^{-4}$ (HPH).

The resulting estimated parameter ranges are reported in Table \ref{tab:4paramFit}.  The corresponding curve fits are not shown as they were visually identical  to those shown in Fig.~\ref{fig:5param}. For both the control and hypertensive animals, the estimated parameter values for $k_{1M}$ varied inversely with $c_M$.  For the hypertensive animals, the geometric parameter ranges  exhibit a small increase in vessel wall inner radius $R_{in}$  and a large increase in reference wall thickness $H$. The ranges of values for the elastic moduli associated with collagen ($k_{1M},k_{2M}$) are substantially higher while the range of values for the modulus associated with elastin ($ c_{M}$) is substantially lower, compared to the control animals.  Furthermore, the range for $c_M$ includes zero, indicating that the hypertensive model can be fit well to the data with a negligible biomechanical contribution of elastin to the stress-strain and wall thickening response to increasing pressure. 

\begin{table*}[tbh!]
\caption{Estimated parameter ranges for the final  model  in both the control (CTL) and hypertensive (HPH) animals based on optimization with 4 parameters. The cost ${\cal J}$ was allowed to increase by no more than 10\% in establishing the estimated parameter ranges.}
\centering
\begin{tabular}{|c||c|c|r||r|r|}
\hline
& Param. & Units & Initial & CTL (\S 3.4) & HPH (\S 3.4)   \\
\hline
$m$ & & & & 4 & 4  \\
\hline
Geom. & $R_{in}$ & $\mu$m & 1000 &  373.27-379.98 & 380.91-397.65 \\
& $H$ & $\mu$m & $T_1^{data}$ & 44.79-46.01 &  61.06-63.31
 \\
\hline
Media & $c_M$ & kPa &   &  19.45-23.50 (fixed) & 0.00-3.10 (fixed)  \\
& $k_{1M}$ & kPa &  1   &0.42-0.87 & 10.59-14.07 \\
& $k_{2M}$ & - & 0.839  & 0.51-0.66 & 1.18-1.34\\
\hline
Adv. & $c_A$ & kPa &   &  $=c_M$ & $=c_M$ \\
& $k_{1A}$ & kPa &  &  $=k_{1M}$ & $=k_{1M}$\\
& $k_{2A}$ & - &  & $=k_{2M}$ & $=k_{2M}$ \\ 
\hline
${\cal J}$ ($\cdot 10^{-4}$) &  & - & &  1.64-1.81 & 0.78-0.86 \\
\hline 
\end{tabular}
\label{tab:4paramFit}
\end{table*}

\section{Discussion and Conclusions} 

This study develops a new mechanistic wall model relating transmural blood pressure and vessel lumen area using a two-layer nonlinear hyperelastic HGO model incorporating residual stresses and anisotropy.  The new model is calibrated using  pressure and wall thickness data from  \cite{TC10} for control and hypertensive mice.  In the hypertensive animals, pulmonary hypertension is induced by placing the animals in a hyperbaric chamber exposing them to hypoxia for 10 days.  Model calibration and systematic model reduction is achieved by combining sensitivity analysis,  subset selection, and parameter estimation.  The results demonstrate that this detailed structural continuum mechanics model, containing a large number of  parameters, can be systematically reduced to capture differences in key model parameters between control and hypertensive animals. 

To our knowledge, this is the first study to carry out robust parameter estimation and model reduction by simultaneously predicting the increase in lumen area and the decrease in wall thickness as pressure is increased.  The results reveal that these biomechanical responses can be accurately captured using a model that retains a single set of three elastic moduli  delineating the contributions of collagen and elastin under the  loading protocol of the associated experiments. Specifically, the material parameter associated with elastin ($c_M$) was the dominant contributor to nonlinear stiffening in the control animals.  By contrast, in the hypertensive animals, the contribution of $c_M$ was negligible and nonlinear stiffening is dominated by material parameters associated with collagen ($k_{1M},k_{2M}$).  Taken together, these findings are consistent with well-known increases in collagen content in the wall of large pulmonary arteries with hypoxia-induced PH \cite{OWT09,WC12}.  Note that our analysis uses the same opening angle value ($\alpha=94.2\degree$~\cite{XPM08}) for both the control and hypertensive models, based on the only known measurements of this quantity in a similar vessel and species after 10-days of hypoxic conditions \cite{HSD01} (see Fig.~7 therein).

The robustness of our model and approach is evidenced by its accurate and simultaneous prediction of both pressure and wall thickness changes under deformation,  for both control and hypertensive data sets.
Our systematic approach to parameter identifiability, subset selection and model reduction decreased the overall number of parameters in the model while preserving the quality of curve-fits to the data at each stage of the iterative procedure. 

Limitations include common parameter estimation challenges when the number of model parameters and/or variables increases relative to the variables or quantities of interest for which data is available, as well as the lack of known nominal values for some model parameters in large pulmonary arteries.  One challenge is non-uniqueness of parameter estimates due to the infeasability of guaranteeing a solution of the optimization problem that is a global minimum of the cost function across the parameter landscape.  A second challenge is the local nature of sensitivity measures underlying the identifiability techniques used in this study, i.e.~the final reduced model is not guaranteed to be unique.  Indeed, the accuracy and robustness of the approaches presented in this study can be enhanced through both extended {\em ex vivo} and {\em in vivo} studies informed by the model presented here. Our approach for sensitivity and identifiability analysis and model reduction is rooted in prior works describing parameter subset selection techniques  \cite{BVV99,CBCL09,PNS21,QM09} using an eigendecomposition of  the matrix $\chi^T \chi$, but similar results could likely be obtained using other methods. 
While global sensitivity analysis techniques exist \cite{KI17,SRA08,M18}, most subset selection techniques are local. The method for identifiability analysis used here is based on eigenvalues but, as discussed in several previous studies, similar results can be obtained using other methods~\cite{H19,MXP11,O13}.  Overall, sensitivities or unidentifiable parameters for particular variables of quantities of interest can  suggest which types of data will be most influential in an expanded data set.  Where practical, examples of extensions include augmentation of {\em ex vivo} biomechanical testing to include measurement of the vessel opening angle, as well as incorporation of {\em in vivo} data measuring BP, flow and lumen area prior to sacrifice of the animal(s). 

Our model is also based on assumptions of hyperelastic deformation and geometric idealization of the stress-free reference state (Fig.~\ref{fig:configs}b)  as a segment of a cylindrical ring; in reality, the vessel wall may exhibit viscoelastic effects under pressurization and/or deviate from circular arcs in the cut open rings.  The new pressure-area relation developed in this study has the potential for incorporation into 1D cardiovascular network models of coupled fluid-solid dynamics in large pulmonary arteries. Some possible approaches  include direct incorporation and coupling of the pressure-area relation within the 1D fluids network solver or, alternatively, using the pressure-area relation as a high fidelity model for emulation using simpler empirical models \cite{LWG84,O99,VBZ11} or statistical models.  Overall, the techniques and findings presented in this study demonstrate the potential for  development and systematic reduction of more realistic models of key relations (e.g.~pressure-area) through the integration of data-driven mathematical approaches for {\em ex vivo} experiments with modeling approaches predicting {\em in vivo} dynamics in cardiovascular biomechanics.

\section{Appendix}
\label{app}
\begin{table*}[tbh!]
\caption{List of all parameters for the model developed in this study. Estimated parameters listed with an asterisk (*) are ultimately fixed (during the course of the model reduction). Note that the model no longer depends on the parameter $L$ when the twist angle $\Phi$ is assumed to be zero (see (\ref{eq:deformation})).  }
\resizebox{\columnwidth}{!}{
\begin{tabular}{|l||l|l|l|l||l|l|}
\hline
Type& Parameter & Description & Units & Role & Fixed & Initial  \\
&&&&& Value & Value\\
\hline
Geometric & $R_{in}$ & Inner radius in $\Omega_0$  & $\mu$m & Estimated &  & 1000 \cite{TC10} \\
 & $R_{out}(=R_{in} + H)$ &  Outer radius in $\Omega_0$ & $\mu$m & Dependent &   & N/A   \\
& $H$ & Vessel wall thickness in $\Omega_0$   &$\mu$m & Estimated &  & $T_1^{data}$\cite{TC10}  \\
& $H_M(=\frac{2}{3}H)$ & Media thickness in $\Omega_0$   &$\mu$m & Dependent &  & N/A  \\
& $\alpha$ & Opening angle in $\Omega_0$ & deg. & Fixed & 94.2  \cite{XPM08} &  \\ 
& $L$ & Axial length in $\Omega_0$ &   $\mu$m & Eliminated & N/A &  \\
& $\lambda_z$ & Axial stretch in deformation & - & Fixed & 1.4 \cite{TC10}  &  \\
& $\Phi$ & Twist angle in deformation & - & Fixed & 0.0 &\\
\hline
Media& $c_M$ &  elastic modulus (iso.) & kPa & Estimated & & 10  \\
& $k_{1M}$ &  elastic modulus (aniso.) & kPa & Estimated &   & 1\\
& $k_{2M}$ &  elastic parameter (aniso.) & - & Estimated &  & 0.839 \cite{HGO00} \\
 & $\beta_M$ & collagen fiber angle & deg. & Fixed & 29 \cite{HGO00} &  \\
\hline
Adventitia & $c_A$ & elastic modulus (iso.) & kPa & Estimated* & & 10  \\
& $k_{1A}$ &  elastic modulus (aniso.) & kPa & Estimated* &  &0.3 \\
& $k_{2A}$ & elastic parameter (aniso.) & - &Estimated* &  & 0.711 \cite{HGO00} \\
& $\beta_A$ & collagen fiber angle & deg. & Fixed & 62 \cite{HGO00} &  \\
\hline 
\end{tabular}
}
\label{tab:appTable}
\end{table*}

\subsection{Complete set of model parameters}
For convenience, the full set of model parameters, their descriptions, units, designation  of parameter type (estimated, fixed, dependent or eliminated) and the associated fixed or initial values are summarized in Table \ref{tab:appTable}.

\subsection{Pressure-area relation}
A Github link to Matlab functions for the integrands $ {\cal F}_M(r_{in},r)$ and 
${\cal F}_A(r_{in},r)$ in (\ref{eq:pAInt}) is available upon request.

\begin{acknowledgements}
 We would like to acknowledge Michelle Bartolo for developing the digital illustration in Fig.~1a.
\end{acknowledgements}


\begin{thebibliography}{}
%
%


\bibitem{AP07}
Azer K, Peskin CS (2007) A one-dimensional model of blood flow in arteries with friction and convection based on the Womersley velocity profile. Cardiovasc Eng 7:51-73

\bibitem{BBG16}
Battista C, Bia D, German YZ, Armentano RL, Haider MA, Olufsen MS (2016) Wave propagation in a 1D fluid dynamics model using pressure-area measurements from ovine arteries. J Mech Med Biol 16:1650007

\bibitem{BVV99}
Burth M, Verghese GC, V\'{e}lez-Reyes M (1999) Subset selection for improved parameter estimates in on-line identification of a synchronous generator.  IEEE Trans Power Syst 14:218-225

\bibitem{CBCL09}
Cintr\'{o}n-Arias A, Banks HT, Capaldi A, Lloyd AL (2009) A sensitivity matrix based methodology for inverse problem formulation. J Inverse Ill-posed  17:545-564

\bibitem{CGL16}
Chen WW, Gao H, Luo XY, Hill NA (2016) Study of cardiovascular function using a coupled left ventricle and systemic circulation model. J Biomech 49:2445-2454

\bibitem{CQR21}
Colebank MJ, Qureshi MU, Rajagopal S, Krasuski RA, Olufsen MS (2021) A multiscale model of vascular function in chronic thromboembolic pulmonary hypertension. Am J Physiol  321:H318-H338

\bibitem{H19}
Haargaard Olsen C, Ottesen JT, Smith RC, Olufsen MS (2019) Parameter subset selection techniques for problems in mathematical biology. Biol Cybern  113:121-138

\bibitem{HGO00}
Holzapfel GA, Gasser TC, Ogden RW (2000) A new constitutive framework for arterial wall mechanics and a comparative study of material models. J Elasticity 61:1-48

\bibitem{HSD01}
Huang W, Sher YP, Delgado-West D, Wu JT, Peck K, Fung YC (2001) Tissue remodeling of rat pulmonary artery in hypoxic breathing. I. Changes of morphology, zero-stress state, and gene expression. Ann Biomed Eng 29:535-551

\bibitem{KI17}
Kucherenko S, Iooss B (2017) Derivative-based global sensitivity measures, in 
Handbook of Uncertainty Quantification, Eds., R Ghanem, D Higdon, and
H Owhadi, 1241-1263, Springer, New York

\bibitem{LWG84}
Langewouters GJ, Wesseling KH, Goedhard WJ (1984) The static elastic properties of 45 human thoracic and 20 abdominal aortas in vitro and the parameters of a new model. J Biomech 17:425-35

\bibitem{MAP07}
Matthys KS, Alastruey J, Peiró J, Khir AW, Segers P, Verdonck PR, Parker KH, Sherwin SJ (2007) Pulse wave propagation in a model human arterial network: assessment of 1-D numerical simulations against in vitro measurements. J Biomech 40:3476-86

\bibitem{M18}
Marquis AD, Arnold A, Dean-Bernhoft C, Carlson BE, Olufsen MS (2018) Practical identifiability and uncertainty quantification of a pulsatile cardiovascular model.  Math Biosci 304:9-24

\bibitem{MAB09}
McLaughlin VV, Archer SL, Badesch DB, Barst RJ, Farber HW, Lindner KR, Mathier MA, McGoon MD,Park MH, Rosenson RS, Rubin LJ, Tapson VF, Varga J (2009) ACCF/AHA 2009 expert consensus document on pulmonary hypertension: A report of the American College of Cardiology Foundation task force on expert consensus documents and the American Heart Association. Circulation 119:2250-2294

\bibitem{MXP11}
Miao H, Xia X, Perelson AS, Wu H (2011) On identifiability of nonlinear ODE models and applications in viral dynamics. SIAM Rev 53:3-39

\bibitem{M04}
Michiels C (2004) Physiological and pathological responses to hypoxia. Am J Pathol 164:1875-1882

\bibitem{NM65}
Nelder J, Mead R (1965) A simplex method for function minimization. Comput J 7:308-313

\bibitem{O99}
Olufsen MS (1999) Structured tree outflow condition for blood flow in larger systemic arteries. Am J Physiol 276:H257– H268

\bibitem{O13}
Olufsen MS, Ottesen JT (2013) A practical approach to parameter estimation applied to model predicting heart rate regulation. J Math Biol 67:39-68

\bibitem{OWT09}
Ooi  CY, Wang Z, Tabima DM, Eickhoff JC, Chesler NC (2010) The role of collagen in extralobar pulmonary artery stiffening in response to hypoxia-induced pulmonary hypertension. Am J Physiol 299:H1823–1831

\bibitem{PCO20}
Paun LM, Colebank MJ, Olufsen MS, Hill NA, Husmeier D (2020) Assessing model mismatch and model selection in a Bayesian uncertainty quantification analysis of a fluid-dynamics model of pulmonary blood circulation. J R Soc Interface 17:20200886

\bibitem{PNS21}
Pearce KJ, Nellenbach K, Smith RC, Brown AC, Haider MA (2021) Modeling and parameter subset selection for fibrin polymerization kinetics with applications to wound healing. Bull Math Biol 83:47

\bibitem{QM09}
Quaiser T, M\"{o}nnigmann M (2009) System identifiability testing for unambiguous mechanistic modeling -- application to JAK-STAT, MAP kinase, and NF-$\kappa$ B signaling pathway models. BMC Systems Biology 3:50

\bibitem{QCP18}
Qureshi MU, Colebank MJ, Paun LM, Ellwein-Fix L, Chesler N, Haider MA, Hill NA, Husmeier D, Olufsen MO (2018) Hemodynamic assessment of pulmonary hypertension in mice: a model based analysis of the disease mechanism. Biomech Model Mechanobiol 18:219-243

\bibitem{QVS14}
Qureshi MU, Vaughan GD, Sainsbury C, Johnson M, Peskin CS, Olufsen MS, Hill NA (2014) Numerical simulation of blood flow and pressure drop in the pulmonary arterial and venous circulation. Biomech Model Mechanobiol 13:1137-54

\bibitem{RH19}
Ramachandra AB, Humphrey JD (2019) Biomechanical characterization of murine pulmonary arteries. J Biomech. 84:18-26

\bibitem{RR08}
Rich S, Rabinovitch M (2008) Diagnosis and treatment of secondary (non-category 1) pulmonary hypertension. Circulation 118:2190-2199

\bibitem{R71}
Rothenberg TJ (1971) Identification in parametric models. Econometrica 39:577-591

\bibitem{SRA08}
Saltelli A, Ratto M, Andres T, Campolongo F, Cariboni J, Gatelli D,
Saisana M, Tarantola S (2008) Global sensitivity analysis: the primer, John
Wiley and Sons, Chichester, UK

\bibitem{S08}
 Shampine LF (2008) Vectorized adaptive quadrature in MATLAB. J Comp Appl Math 211:131-140

 \bibitem{SMC19}
Simonneau G, Montani D, Celermajer DS, Denton CP, Gatzoulis MA, Krowka M, Williams PG, Souza R (2019) Haemodynamic definitions and updated clinical classification of pulmonary hypertension. Eur Respir J 53:1801913

\bibitem{TC10}
Tabima DM, Chesler NC (2010) The effects of vasoactivity and hypoxic pulmonary hypertension on extalobar pulmonary arterial biomechanics. J  Biomech 43:1864-1869

\bibitem{VBZ11}
Valdez-Jasso D, Bia D, Zócalo Y, Armentano RL, Haider MA, Olufsen MS (2011) Linear and nonlinear viscoelastic modeling of aorta and carotid pressure-area dynamics under in vivo and ex vivo conditions. Ann Biomed Eng 39:1438-1456.

\bibitem{VS11}
Van de Vosse FN, Stergiopulos N (2011) Pulse wave propagation in the arterial tree, Ann Rev Fluid Mech 43:467-499. 

\bibitem{WC12}
Wang Z, Chesler NC (2012) Role of collagen content and cross-linking in large pulmonary arterial stiffening after chronic hypoxia. Biomech Model Mechanobiol, 11:279–289

\bibitem{WFM95}
Wohrley JD, Frid MG, Moiseeva EP, Orton EC, Belknap JK, Stenmark KR (1995) Hypoxia selectively induces proliferation in a specific subpopulation of smooth muscle cells in the bovine neonatal pulmonary arterial media. J Clin Invest 96:273-281

\bibitem{XPM08}
Xu M, Platoshyn O, Makino A, Dillmann WH, Akassoglou K, Remillard CV, Yuan JXJ (2008) Characterization of agonist-induced vasoconstriction in mouse
pulmonary artery. Am J Physiol  294: H220-H228

\bibitem{ZMZ18}
Zambrano BA, McLean NA, Zhao X, Tan JL, Zhong L, Figueroa CA, Lee LC, Baek S (2018) Image-based computational assessment of vascular wall mechanics and hemodynamics in pulmonary arterial hypertension patients. J Biomech 68:84-92

\end{thebibliography}


\end{document}